\documentclass[%
 reprint,
showpacs,preprintnumbers,
 amsmath,amssymb,
 aps,
]{revtex4-1}

\usepackage{graphicx}% Include figure files
\usepackage{dcolumn}% Align table columns on decimal point
\usepackage{bm}% bold math
\usepackage{hyperref}% add hypertext capabilities
\newcommand{\ket}[1]{\left|#1\right\rangle}
\begin{document}

\title{A simple method for generating Bose-Einstein condensates in a weak hybrid trap}
\author{M. Zaiser}
\author{J. Hartwig}
\author{D. Schlippert}
\author{U. Velte}
\author{N. Winter}
\author{V. Lebedev}
\author{W. Ertmer}
\author{E. M. Rasel}
\email{Rasel@iqo.uni-hannover.de}
\affiliation{%
 Institute of Quantum Optics and\\ Centre for Quantum Engineering and Space-Time Research - QUEST, Cluster of Excellence, Leibniz Universit\"at Hannover, Welfengarten 1, 30167 Hannover, Germany
}%

\date{\today}% It is always \today, today,

\begin{abstract}
We report on a simple novel trapping scheme for the generation of Bose-Einstein condensates of $^{87}$Rb atoms. This scheme employs a near-infrared single beam optical dipole trap combined with a weak magnetic quadrupole field as used for magneto-optical trapping to enhance the confinement in axial direction. Efficient forced evaporative cooling to the phase transition is achieved in this weak hybrid trap via reduction of the laser intensity of the optical dipole trap at constant magnetic field gradient.
\end{abstract}

\pacs{37.10.Jk, 67.85.Hj, 67.85.Jk}% PACS, the Physics and Astronomy
\maketitle

A simple and robust method for the generation of quantum degenerate atomic gases with decent particle number and repetition rate is of interest for the study of their fundamental properties~\cite{Greiner02Nature,Billy08Nature,*Roati08Nature,Scherer10PRL} or their applications in atomic inertial sensors~\cite{Canuel06PRL} and gravimeters~\cite{Peters99Nature}, as well as in microgravity~\cite{Zoest10Science}. Here, we describe a very simple method for the generation of a Bose-Einstein condensate (BEC) in a single beam near-infrared optical dipole trap (ODT). Optical dipole traps offer great potential with respect to the criteria mentioned above. Forced evaporative cooling in such traps is usually achieved by reducing the power of the ODT laser beam~\cite{Adams95PRL} and rethermalization times are generally short due to the high trapping frequencies in the kHz-regime usually provided by an ODT. However, power reduction also reduces the confinement of the atoms in the trap which in turn negatively affects trap frequencies, peak atomic density, elastic collision rate, and as a consequence the efficiency of forced evaporation. This counteracts the gain in phase space density by the cooling of the atomic cloud, thus preventing the regime of run-away evaporation in an ODT with this simple method~\cite{O'Hara01PRAR}. Quantum degeneracy can nonetheless be reached in these traps, provided the initial atomic and phase space densities were high enough~\cite{Barrett01PRL,Granade02PRL}. Nevertheless, the necessary compromise between high initial densities and high initial trapping volume severely affects the maximum number of particles in the BEC and additional sophisticated concepts for reaching optimized initial atomic and phase space densities may be required~\cite{Winoto99PRAR,*Han00PRL}.

The most simple realization of an ODT is to focus one single far-off resonant high power laser beam onto the atomic ensemble. However, due to the rather low confinement of the atoms in the \emph{axial} direction of these single beam ODTs, the high initial atomic and phase space densities needed to reach quantum degeneracy are very hard to realize. This is particularly the case for single beam ODTs formed from a near-infrared laser source~\cite{Rayleigh}. In contrast, the wavelength of a CO$_2$ laser of $\sim10.6$~${\rm {\mu}}$m provides an axial trapping frequency an order of magnitude higher compared to an ODT formed from e.g. a Nd:YAG laser at a wavelength of 1064~nm. This stronger confinement is still enough to allow for the realization of a BEC with more than $10^5$ atoms~\cite{Gericke07APB}. Nevertheless, this method is bound to the use of far-infrared laser wavelengths with all the associated technical implications. If the use of laser wavelengths in the near-infrared is desired, additional, more sophisticated approaches are required. One way to combine high initial trapping volume during the loading of the ODT with the high trapping frequencies needed for fast evaporative cooling is a compressible ODT employing a zoom lens technique~\cite{Kinoshita05PRAR,Chang05NatPhys,*Chang06}. An alternative scheme for BEC generation is the use of a strong hybrid trap, i.e. the combination of a near-infrared single beam ODT with preceding rf-evaporation in a strong magnetic quadrupole trap~\cite{Lin09PRA}.

\begin{figure*}[tb]
\includegraphics{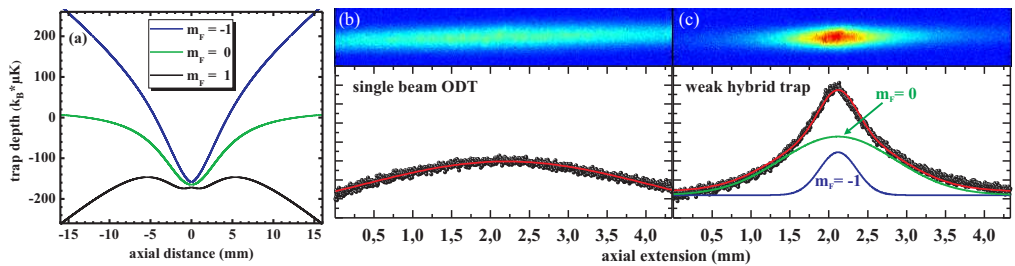}
\caption{\label{fig:hybrid} (Color online) (a) Exemplary potentials of the weak hybrid trap for the atoms in the three different $m_F$-substates as a function of the distance in axial direction to the minimum beam radius $w_0=53$~${\rm {\mu}}$m of the ODT. The minimum trap depth is U$_0=k_B\times190$~${\rm {\mu}}$K and the potentials were calculated for a gradient of the magnetic quadrupole field of 10~G/cm and a spatial offset between the centers of ODT and quadrupole field of 200~${\rm {\mu}}$m. Absorption images of a cloud of $2\times10^6$~atoms (b) in the single beam ODT and (c) in the weak hybrid trap after 1~ms of free expansion reveal the corresponding density profiles. The enhancement of the atomic density in the weak hybrid trap is clearly visible. Furthermore, the density distribution of the atomic cloud in the weak hybrid trap (c) is composed out of two Gaussian density distributions, where the narrower distribution (blue) represents atoms in $\ket{m_F=-1}$ experiencing the additional magnetic confinement. On the other hand, atoms in the magnetic-field-insensitive spin state $\ket{m_F=0}$ (green) only experience the ODT potential and thus show a broader density distribution. Atoms in $\ket{m_F=+1}$ are repelled from the auxiliary magnetic field, as shown in (a).}
\end{figure*}

In this paper, we present a simple method to enhance the peak density of the atomic cloud in a single beam ODT deduced from a near-infrared laser. Furthermore, this approach also allows to prevent the most severe reduction of trap frequency in the axial direction of the ODT during forced evaporative cooling, thus actually allowing an \emph{increase} in atomic density. It relies on the use of an additional, rather weak auxiliary magnetic quadrupole field which by itself is too low to levitate the atoms against gravity. Due to this additional magnetic field, the trapping potential becomes dependent of the specific $\ket{m_F}$-substate of the atom. Therefore, $^{87}$Rb atoms in $\ket{F=1,m_F=-1}$ ($\ket{F=1,m_F=+1}$) experience an additional attractive (repulsive) force, as is shown schematically in Fig.~\ref{fig:hybrid} (a). Making use of this method, we demonstrate the generation of BEC in a single beam ODT at a wavelength of 1960~nm with a volume rather large compared to similar experiments~\cite{Barrett01PRL,Gericke07APB,Chang05NatPhys,*Chang06,Lin09PRA}.

The ODT used here is derived from a single mode Thulium-doped fiber laser (TLR-50-1960-LP, IPG Photonics) operating at a wavelength of 1960~nm with $M^2<1.1$ and a FWHM linewidth of 1~nm. This laser beam is focused onto the atoms with a minimum $1/e^2$ beam radius of $w_0=53$~${\rm {\mu}}$m. For a maximum laser power of 10~W used in the experiments presented here, this corresponds to a maximum trap depth of U$_0=k_B\times190$~${\rm {\mu}}$K, a radial (axial) trapping frequency of $\omega_{\text{rad}}=2\pi\times800$~Hz ($\omega_{\text{ax}}=2\pi\times7$~Hz), and a maximum differential AC Stark shift of -17~MHz between the two states $5^2$S$_{1/2}$ and $5^2$P$_{3/2}$ used for laser cooling. The laser beam propagates in horizontal direction with an angle of $25^\circ$ with respect to the axes of the anti-Helmholtz coils for operating the three-dimensional magneto-optical trap (3D MOT).

The auxiliary magnetic trapping field of the weak hybrid trap is very conveniently provided by the two coils also used to generate the magnetic quadrupole field for operating the 3D MOT. Fig.~\ref{fig:hybrid} (b) and (c) clearly illustrate the substantial increase in initial peak density of the atomic cloud in the ODT by more than a factor of 2 due to this auxiliary trapping field. The two absorption images show clouds of $2\times10^6$~atoms (b) in the single beam ODT and (c) in the weak hybrid trap after a 1~ms time-of-flight (TOF) expansion. A closer look at the density profile of the atomic ensemble in the weak hybrid trap reveals a form consisting of two Gaussians which we interpret as clouds in the two spin states $\ket{m_F=0}$ (green) and $\ket{m_F=-1}$ (blue). Atoms in the state $\ket{m_F=+1}$ are repelled from the auxiliary magnetic field, as shown in Fig.~\ref{fig:hybrid} (a).

For the experimental studies presented in this paper, the 3D MOT is first loaded from the cold atomic beam of a 2D MOT for one second with $5\times10^8$~atoms with the ODT already being switched on at maximum power. The 2D MOT is then switched off and the 3D MOT is tuned to a temporal dark MOT phase~\cite{Townsend96PRA} at constant magnetic field gradient for 55~ms to enhance the density of the atomic cloud and thus optimize the loading of the ODT~\cite{Kuppens00PRA}. During this phase, the detuning of the MOT cooling light is changed to $\delta_{\text{MOT}}=-25\,\gamma$ ($\gamma=6.07$~MHz) and the total repump intensity is reduced to 4~${\rm {\mu}}$W/cm$^2$. This results in a population of the lower hyperfine ground state $5^2$S$_{1/2}\ket{F=1}$ of about 85~\%. To prevent atom loss in the ODT due to hyperfine changing collisions, all the atoms are pumped into the $5^2$S$_{1/2}\ket{F=1, m_F}$-manifold at the end of this phase by completely shutting off the repumping light 600~${\rm {\mu}}$s before the MOT cooling light.

\begin{figure}[b]
\includegraphics{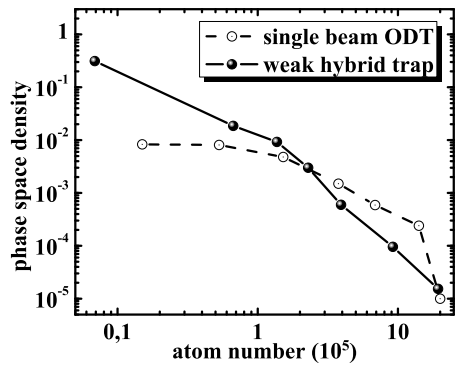}
\caption{\label{fig:EvapComp} Comparison of the evolution of phase space density in the single beam ODT (open circles) and in the weak hybrid trap (solid circles) during forced evaporative cooling. The power of the ODT laser beam is reduced using a Pockels cell and a subsequent Glan-Laser-Polarizer. Lines are guides to the eye.}
\end{figure}

With this loading procedure, $2\times10^6$~atoms are trapped in the single beam ODT at an initial temperature of 35~${\rm {\mu}}$K and a corresponding initial phase space density of about $1-2\times 10^{-5}$. This is in good agreement with the values found experimentally for the temporal dark MOT without ODT. Evaporative cooling in this system is achieved via simple reduction of the ODT laser power by means of a Pockels cell and a subsequent Glan-Laser-Polarizer in the beam path of the ODT. The power is reduced in six independent successive linear ramps, where the slope of each ramp was optimized to the biggest gain in phase space density of the atomic ensemble. To compare evaporative cooling in the single beam ODT and in the weak hybrid trap, respectively, the magnetic quadrupole field of the 3D MOT with a gradient of 10~G/cm is switched off or simply kept on during this evaporation phase.

As expected, the peak density and thus the elastic collision rate of the atomic ensemble in the single beam ODT continuously decrease during forced evaporative cooling due to the reduction of trapping frequencies. This leads to a continuous decrease of the rate of power reduction during evaporation. The optimum power ramp $P(t)$ hence takes the form~\cite{O'Hara01PRAR}
\begin{equation}
P(t)=P_0(1+t/\tau)^{-\beta}\, ,
\end{equation}
with the constants $\tau$ and $\beta$ and the power $P_0$ of the ODT laser beam prior to evaporative cooling. The total duration of the six successive linear ramps is about 40~s and the minimum achievable temperature is 200~nK, where the temperature has been determined using TOF measurements. However, due to the continuous reduction of atomic density, the efficiency of evaporation, i.e. the relative increase in phase space density with decreasing atom number, declines and eventually vanishes, as shown in Fig.~\ref{fig:EvapComp}. The maximum reachable phase space density in the single beam ODT inferred from temperature, atom number, and calculated corresponding trapping frequencies after each linear evaporation ramp, is hence limited to $\sim 1\times 10^{-2}$.

\begin{figure}[t]
\includegraphics{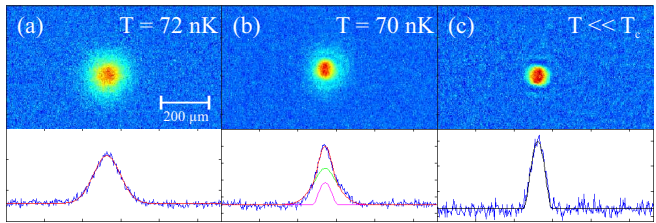}
\caption{\label{fig:bimodal} (Color online) Absorption images and corresponding density profiles of atomic clouds at different temperatures released from the weak hybrid trap. The time-of-flight for each picture is 21~ms and the phase transition from a thermal gas in (a) to a bimodal density distribution in (b) to a quasi pure BEC in (c) is clearly visible. The quasi pure BEC consists of $10^4$~atoms.}
\end{figure}

In contrast, the auxiliary magnetic trapping field of the weak hybrid trap not only increases the \emph{initial} peak density of the atomic cloud in the ODT, it also provides a constant confinement of the atoms in the $\ket{m_F=-1}$-state along the axis of the ODT. This allows for a constant \emph{increase} in peak atomic density by an order of magnitude during evaporation and consequently a reduction of the time needed for the evaporation phase by a factor of 2 compared to the single beam ODT. Additionally, the minimum temperature of the atomic cloud of $\sim 75$~nK observed with this method is substantially lower in comparison with the single beam ODT. Consequently, the phase space density of the atomic ensemble inferred from TOF-studies grows continuously after each intensity ramp up to a maximum phase space density on the order of $3\times 10^{-1}$ (see Fig.~\ref{fig:EvapComp}).

The phase transition to BEC is eventually observed by lowering the output power of the fiber laser itself during the last evaporation ramp in addition to the use of the Pockels cell. This approach is necessary due to constraints in the minimum power reachable using only the polarization optics. The onset of quantum degeneracy is clearly revealed by the bimodal density distribution of the atomic cloud shown in Fig.~\ref{fig:bimodal}(b). Reducing the ODT laser power even further allows the generation of quasi pure condensates (see Fig.~\ref{fig:bimodal}(c)) with more than $1\times 10^4$ atoms every 20~seconds.

In conclusion, we have demonstrated a very simple and robust method to increase the peak atomic density in single beam near-infrared optical dipole traps. In addition, this method also helps to prevent the severe decrease of trapping frequency in the axial direction of such a trap during forced evaporative cooling. The method relies on the application of an auxiliary magnetic trapping field during forced evaporation, for which we simply used the two anti-Helmholtz coils for operating the 3D MOT. Employing this weak hybrid trap allowed us to generate Bose-Einstein condensates with more than $1\times 10^4$ atoms in a single beam optical dipole trap at a wavelength of 1960~nm. In the case of $^{87}$Rb, the use of this wavelength is compatible with near-resonant laser cooling to increase the initial phase space density prior to forced evaporative cooling. This is due to the negative, but moderate differential AC Stark shift, just like in the case of an ODT formed by a CO$_2$ laser beam~\cite{Friebel98APB}. Laser cooling in combination with employing a higher magnetic field gradient during the evaporation phase and optically pumping all the atoms into the low-field seeking spin states $\ket{F=1, m_F=-1}$ or $\ket{F=2, m_F=+2}$ provides additional potential for further improving the efficiency of evaporation and thus the particle number in the BEC and the repetition rate.

We acknowledge T. W\"{u}bbena for valuable contributions during the early stages of the experiment and C. Klempt, O. Topic, and M. Riedmann for fruitful discussions. This work is supported by the European Union (FINAQS), the Deutsche Zentrum f\"{u}r Luft- und Raumfahrt DLR (PRIMUS), the European Science Foundation (EuroQUASAR), and the Centre for Quantum Engineering and Space-Time Research QUEST. J.H. and D.S. would like to acknowledge financial support from DLR and HALOSTAR, respectively.

\bibliography{hybrid}

\begin{thebibliography}{10}
\bibitem{Rayleigh}%
  \BibitemOpen
  \bibinfo {note} {The confinement and thus the trapping frequency in the axial
  direction of a single focused laser beam are proportional to the Rayleigh
  length $z_R=\pi w_0^2/\lambda$, where $\lambda$ is the wavelength of the
  laser beam and $w_0$ denotes its minimum radius at the position of the
  focus.}%
  \bibAnnoteFile{Stop}{Rayleigh}%
\end{thebibliography}

\end{document}